\documentclass[a4paper,10pt]{article}
\usepackage{amssymb}
\usepackage{geometry}
\usepackage[usenames,dvipsnames]{color}
\usepackage{slashed}
\usepackage[table]{xcolor}
\usepackage{graphicx}
\usepackage{mathtools}
\usepackage{cite} 

\usepackage{amsthm}
\usepackage{bm}

\theoremstyle{plain}

\theoremstyle{definition}

\theoremstyle{remark}

\title{\bf{Entropic alternatives to initialization}}
\date{}
\author{Daniele Musso\footnote{daniele.musso@usc.es, mudaniele@yahoo.com}}

\begin{document}

\maketitle
\vspace{-20pt}
\begin{center}\it{
Centro de Supercomputaci\'on de Galicia (CESGA),\\
s/n, Avenida de Vigo, 15705 , Santiago de Compostela, Spain}
\end{center}
\vspace{5pt}

\begin{abstract}
Local entropic loss functions provide a versatile framework to define architecture-aware regularization procedures. 
Besides the possibility of being anisotropic in the synaptic space, the local entropic smoothening of the loss function 
can vary during training, thus yielding a tunable model complexity. A scoping protocol where the regularization is strong 
in the early-stage of the training and then fades progressively away constitutes an alternative to standard initialization 
procedures for deep convolutional neural networks, nonetheless, it has wider applicability. We analyze anisotropic, local 
entropic smoothenings in the language of statistical physics and information theory, providing insight into both their 
interpretation and workings. We comment some aspects related to the physics of renormalization and the spacetime structure of 
convolutional networks.

\end{abstract}
\newpage
\tableofcontents

\section{Introduction}

Insight and methods coming from physics have ever since helped to improve our understanding and design of neural networks. The interdisciplinary potential of techniques borrowed from statistical physics and information theory, such as entropic regularizations and renormalization, has not yet exhausted its drive in machine learning.  

Stochastic gradient descent (SGD) proved to be a particularly suited optimization algorithm to train deep neural networks. 
This is mainly due to the properties of its noise, which is related to the Hessian matrix of the loss function. 
Efficient escaping from sharp relative minima is an example of a useful effect directly descending from the characteristics of the SGD noise. 
Besides, it is argued that regularization effects due to noise bias the stochastic gradient descent algorithm towards encountering, 
both systematically and efficiently, solutions belonging to clusters characterized by a high value of the test accuracy.
If not universal, this phenomenon is believed to be generic for networks working in a regime well below their critical 
capacity. 

To the purpose of understanding the virtues of SGD algorithms from a theoretical viewpoint, and in order to define approaches able to enhance them, 
it has been recently proposed to modify the loss function by means of a local solution-counting term, a \emph{local entropy}. 
Local entropy represents a refinement of standard entropy and defines a coarse-graining technique helpful in understanding and improving deep neural networks.
In particular, it offers a tunable way of encouraging the training towards clusterized solutions through a smoothened version of the original loss function.
More precisely, the modified loss function allows us to perform a large-deviation analysis, biasing the statistical measure away 
from Gibbs typicality towards regions with a high density of high-accuracy weight configurations \cite{Baldassi_2015,Baldassi_2016aa}.

The local entropy framework is attractive in many respects: the relation to statistical mechanics improves its interpretability; 
it admits time-dependent and anisotropic generalizations, making it a flexible framework allowing for adaptive strategies; 
it  demonstrated a potential in image-classification experiments and constraint-satisfaction problems.
Although being trained with a regularized loss function inspired by statistical mechanics, 
the deep networks used in practice are in general very far from being amenable to an
analytical description. They, in fact, depart in many ways from the regimes where analytical approaches 
can be available. 

Local entropy is associated to a position-dependent Helmholtz free energy defined by convoluting the Boltzmann weight with a specified kernel.
The term \emph{local} refers to the fact that the convolution kernel is significantly different from zero only over a compact region.%
\footnote{Spatial locality in a high-dimensional space like the synaptic space can be a somewhat misguiding concept. 
Indeed, recall that -in a high-dimensional space- the volume of a sphere is sharply concentrated close to its surface. 
A random sampling of such high-dimensional sphere, when uniform in volume, would thereby concentrate in the near-surface region.}
A natural example is provided by
\begin{equation}\label{loc_ent}
 e^{-\beta {\cal F}(\boldsymbol W)} = \sqrt{\frac{\beta \gamma}{2\pi}}\ \int d\boldsymbol W'  e^{-\beta\left[{\cal L}(\boldsymbol W') + \frac{\gamma}{2} \|\boldsymbol W - \boldsymbol W'\|_2^2\right]}\ ,
\end{equation}
where the original loss function ${\cal L}$ plays the role of the energy and $\beta$ represents an inverse temperature. 
With $\boldsymbol W$ we denoted the position in the synaptic space (\emph{i.e.} a vector collecting the weights of the network, 
thus representing its state), $\|\|_2$ is the Euclid-Frobenius norm and $\gamma$ is inversely related to the width of the Gaussian kernel
\begin{equation}
 K(\boldsymbol W, \boldsymbol W') = 
 e^{-\beta \frac{\gamma}{2} \|\boldsymbol W - \boldsymbol W'\|_2^2}\ .
\end{equation}
The idea is to use ${\cal F}(\boldsymbol W)$ as a regularized version of the original loss, where the smoothening scale is controlled by $\gamma^{-\frac{1}{2}}$.

Even before considering more generic kernels $K$, \eqref{loc_ent} can be extended naturally in two ways: 
on the one hand, one can consider anisotropic Gaussian kernels where $\gamma$ takes a different value for different subspaces in the synaptic space. 
This corresponds to weighting differently the distance between two points depending on the direction of their separation and has been dubbed \emph{partial local entropy} \cite{Musso:2020qmt}. 
On the other hand, one can consider time-dependent choices where $\gamma$ follows either a pre-defined schedule or an adaptive protocol. 

The present work is concerned with such generalizations, devoting particular attention to the time-dependent extension of \eqref{loc_ent}, 
which -as we will show- provides a useful alternative to sophisticated initialization procedures in image-classification tasks 
performed with deep convolutional networks. Furthermore, a varying $\gamma$ admits practically useful insight coming from complexity 
theory and the physics of renormalization.

\section{Architecture-aware entropic regularization}

The main point of considering a different $\gamma$ for different directions in the synaptic space, 
namely considering an anisotropic local entropy, consists in treating distinct weights in a different manner.
Since deep neural networks are by construction hierarchical systems, where the nature of the hierarchy is 
connected to combinatorial complexity, the anisotropic extension of \eqref{loc_ent} is theoretically natural. Said otherwise, it seems in general not adequate to treat all the weights on the same footing as far as regularization is concerned.

Assigning different values of $\gamma$  to different weight subspaces (\emph{e.g.} 
a different $\gamma$ for each layer) corresponds to defining a regularization strategy which is adapted to the network architecture. 
It is therefore fair to expect that a suitable, anisotropically tuned, $\gamma$ can permit a better exploitation of the biases intrinsically hard-coded in the architecture itself.

These comments appear particularly suited to deep networks, where the depth has a crucial and transparent role. Nonetheless, the idea can be generalized to other contexts. In general, it amounts to having a tunable neural sensitivity and -as such- it can be connected to studies on neural plasticity.
More theoretically, it is interesting to use the convolutional kernels introduced through \eqref{loc_ent} as a probe to explore (and modify) the local capacity and sensitivity of the network.%
\footnote{More comments on the relation between the kernel size and measures of network complexity are given in Section \ref{discu}.}

\begin{figure}
 \begin{center}
  \includegraphics[width=.69\textwidth]{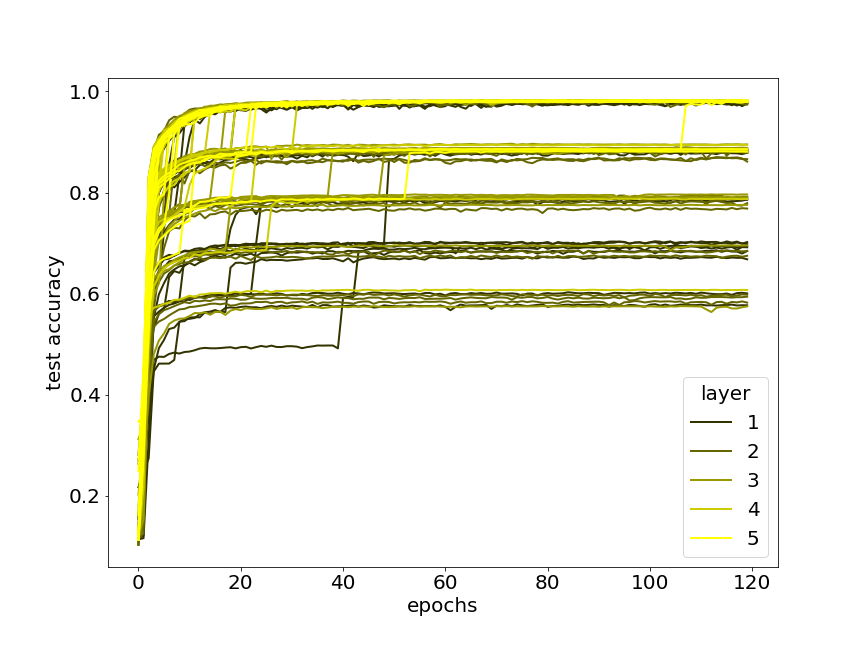}
 \end{center}
 \caption{Single-layer entropic regularization is more effective when applied to progressively deeper layers.}
 \label{anis}
\end{figure}

The analysis of \cite{Musso:2020qmt} considered anisotropic entropic regularization where the anisotropy in the synaptic space
respects the layer structure of the network. That is, all the neurons belonging to a same layer are smoothened over in the same way.
In this sense, the entropic regularization can be architecture-aware.
As a special case, one can consider entropic regularization on a single layer at a time. The numerical experiments described in \cite{Musso:2020qmt}
hinted to the possibly generic fact that single-layer regularization is more effective than multi-layer regularization, on the one side,
and increasingly more effective when applied to increasingly deeper layers, on the other.%
\footnote{The situation can actually be more complicated than just stated, and -for instance- it may depend on the radius of the vicinity over which one smoothens the loss function; we refer to \cite{Musso:2020qmt} for a more detailed discussion. The last output layer, having 
as many neurons as the classes of the task, is actually excluded from the argument. Thus, the ``deepest'' layer is the second-last from the input.}
We corroborate this observations with a new series of experiments performed on MNIST with a 5-layer fully-connected cylindrical network
where all the layers, except the output layer, have 784 neurons. For the training we have considered momentum $\mu=0.9$, constant learning rate $\eta=10^{-4}$,
constant batch-size of 256 images, ReLU activations and no weight-decay regularization. The results are reported in Figure \ref{anis},
which calls for some comments. First, the asymptotic test-accuracy appear to define a ``discrete spectrum''. This reinforces the idea that 
the SGD training encounters solutions belonging to few clusters, representing rare deviations from typicality, yet systematically found.
Considering the same regularization intensity on progressively deeper layers enhances the performance, indicating that the entropic 
regularization is more effective when applied to the synapses corresponding to more complex features. This seems to agree with the intuitive 
idea that the same level of noise is more harmful when affecting a deep rather than a shallow layer.

\section{Information theoretic interpretation of local entropies}

It is useful to study the connections of local entropy to information theory. 
This allows us to appreciate how the modified loss function (\emph{i.e.} the local free energy) is associated to an entropy encoding the similarity of the convolutional kernel and the modified Boltzmann weight. More precisely, the local entropy arises from a relative entropy.

Let us first introduce a modified (local) partition function as
\begin{equation}\label{part}
 \tilde Z(\boldsymbol W) = \int d\boldsymbol W' e^{-\beta {\cal L}(\boldsymbol W')} K(\boldsymbol W,\boldsymbol W')\ .
\end{equation}
For the moment being, we leave the convolutional kernel $K$ generic, yet we assume that it satisfies
\begin{equation}
 K(\boldsymbol W,\boldsymbol W') \geq 0\ , \qquad \text{for all}\qquad \boldsymbol W\ ,\ \boldsymbol W'\ ,
\end{equation}
and
\begin{equation}
 \int d\boldsymbol W'\, K(\boldsymbol W',\boldsymbol W) =
 \int d\boldsymbol W'\, K(\boldsymbol W,\boldsymbol W') =
 1\ , \qquad \text{for all}\qquad \boldsymbol W\ .
\end{equation}
This corresponds to a normalization property%
\footnote{For the specific case of \eqref{loc_ent} the normalization was encoded in the $\sqrt{\frac{\beta\gamma}{2\pi}}$ pre-factor.} thanks to which we have
\begin{equation}
 \int d\boldsymbol W \tilde Z(\boldsymbol W) = \int d\boldsymbol W' e^{-\beta {\cal L}(\boldsymbol W')} \int d\boldsymbol W K(\boldsymbol W,\boldsymbol W')
 = \int d\boldsymbol W' e^{-\beta {\cal L}(\boldsymbol W')} 
 = Z\ ,
\end{equation}
where $Z$ is the standard partition function. Thus, $\tilde Z(\boldsymbol W)$ represents the local contribution to $Z$.%
\footnote{A tilde is adopted throughout the paper to represent the local generalization of the tilded quantity.}

As customary in statistical mechanics, the free energy is derived from the partition function through
\begin{equation}\label{loc_free_ene}
 \tilde F(\boldsymbol W) = -k T\ \ln \tilde Z(\boldsymbol W)\ ,
\end{equation}
which provides a local generalization of the standard derivation. 
We have introduced a Boltzmann constant $k$ and the temperature $T$ (such that $\beta = \frac{1}{KT}$) 
to maintain the connection to statistical physics explicit, yet the experiments will be be performed taking $\beta = 1$.

The partition function \eqref{part} allows us to define the following probability distribution
\begin{equation}\label{dist}
 \tilde \rho(\boldsymbol W,\boldsymbol W') = \frac{e^{-\beta {\cal L}(\boldsymbol W')}}{\tilde Z(\boldsymbol W)} K(\boldsymbol W,\boldsymbol W')\ .
\end{equation}
Note that the probability distribution is to be thought of as $\tilde \rho_{\boldsymbol W}(\boldsymbol W')$, that is, a probability distribution over the synaptic space spanned by $\boldsymbol W'$ where $\boldsymbol W$ plays the role of a parameter (namely, the ``center'' about which we define local quantities).

From the definition of the partition function \eqref{part}, we have the correct normalization of $\tilde \rho$, namely
\begin{equation}
 \int d\boldsymbol W' \tilde \rho(\boldsymbol W,\boldsymbol W') = 1\ .
\end{equation}
A re-arrangement of the terms in \eqref{dist} gives us
\begin{equation}
 \beta {\cal L}(\boldsymbol W') = -\ln \tilde \rho(\boldsymbol W,\boldsymbol W') - \ln \tilde Z(\boldsymbol W) + \ln K(\boldsymbol W,\boldsymbol W')\ ,
\end{equation}
which will be useful shortly.

Now, we follow the standard steps to define the entropy, yet we start from the local free energy \eqref{loc_free_ene}, namely 
\begin{align}\label{ent}
 &\tilde S(\boldsymbol W) = - \frac{\partial \tilde F(\boldsymbol W)}{\partial T} \\ \nonumber
 &= k\left[ \ln \tilde Z(\boldsymbol W)
 + \frac{\beta}{\tilde Z(\boldsymbol W)}  \int d\boldsymbol W' {\cal L}(\boldsymbol W') e^{-\beta {\cal L}(\boldsymbol W')} K(\boldsymbol W,\boldsymbol W')\right] \\ \nonumber
 &= k\left[ \ln \tilde Z(\boldsymbol W)
 - \int d\boldsymbol W' \tilde \rho(\boldsymbol W,\boldsymbol W') \ln \tilde \rho(\boldsymbol W,\boldsymbol W')\right. \\ \nonumber
 &\qquad \qquad \left.
 - \int d\boldsymbol W' \tilde \rho(\boldsymbol W,\boldsymbol W') \ln \tilde Z(\boldsymbol W) 
 + \int d\boldsymbol W' \tilde \rho(\boldsymbol W,\boldsymbol W') \ln K(\boldsymbol W,\boldsymbol W')\right]\\ \nonumber
 &= -k\int d\boldsymbol W' \tilde \rho(\boldsymbol W,\boldsymbol W') \left[\ln \tilde \rho(\boldsymbol W,\boldsymbol W') - \ln K(\boldsymbol W,\boldsymbol W')\right]\\ \nonumber
 &= -k\, D_{\text{KL}} \left[\tilde \rho(\boldsymbol W,\boldsymbol W')\big|\big|K(\boldsymbol W,\boldsymbol W')\right]
 \ .
\end{align}

Equation \eqref{ent} shows that the local entropy $\tilde S(\boldsymbol W)$ corresponds to the relative entropy
among the probability distribution \eqref{dist} and the kernel $K$ \cite{Witten_2020,cover2012elements}.
To gain intuition, equation \eqref{dist} implies that a flat loss ${\cal L}(\boldsymbol W)$ maximizes $\tilde S(\boldsymbol W)$.
Note that, being the kernel $K$ present in the definition of the distribution $\tilde \rho$, 
the relative entropy \eqref{ent} encodes mainly an intrinsic property of ${\cal L}(\boldsymbol W)$ 
rather than a property induced by the shape of the kernel. Yet, the kernel is determining
the region in synaptic space upon which the entropic comparison is made.%
\footnote{Abusing the language to the sake of conveying an intuitive idea, one could say that 
local entropy is an entropy with a \emph{receptive field}, this latter being determined
by the support of the considered convolutional kernel $K$.}

\section{Entropic regularization and tunable complexity}
\label{scoping}

The number of data points introduces a natural resolution scale into the synaptic space \cite{berezniuk2020scaledependent}. 
To rephrase, the complexity of the dataset translates into the complexity of the landscape of the loss function built upon the dataset itself. 
In this perspective, a regularization techniques which smoothens the loss function reduces the complexity of the landscape. 
This can be precisely stated in terms of the Fisher information matrix \cite{berezniuk2020scaledependent}. 
Indeed, a smoothened loss will have --by construction--
a softer dependence on the weights, leading therefore to smaller eigenvalues of the Fisher information matrix.%
\footnote{We will further comment this point in relation to local scale invariance in Subsection \ref{ska}.
For the definition of the Fisher information matrix and its relevance to the present discussion we refer to \cite{cover2012elements}
and \cite{berezniuk2020scaledependent}, respectively.}

The possibility of tuning by hand the complexity of the synaptic space is attractive both on a theoretical and on a practical level. 
Roughly, it corresponds to having a tunable sensitivity for the synapses, which can be exploited to define improved training protocols. 
In practice, this amounts to consider an entropic regularization like \eqref{loc_ent} where the regularization parameter(s) $\gamma$ 
evolves in time during training. Its evolution can either be ruled by a pre-determined schedule or be adaptive. The former possibility 
corresponds to a planned \emph{scoping} of the entropic regularization, the latter introduces an extra intrinsic dynamical ingredient.

The fact that scoping the local entropic regularization can be a good idea is directly suggested by experiments where no scoping is considered. 
In fact, when considering sufficiently complicated image-classification tasks like CIFAR10 or STL10, 
it appears that the entropic regularization provides an advantage in training performance only when combined with an early-stopping protocol \cite{Musso:2020qmt}. 
In other words, the local entropic regularization seems helpful, but only in an early stage of the training process.

Such phenomenon can be understood as follows. The early stage of a stochastic gradient descent is typically noisier than later stages. 
Thereby, filtering away some noise in the initial phase produces a stabler and more effective training signal.%
\footnote{This can be put in analogy to the effect of momentum, we discuss this comparison in Subsection \ref{deno}.
We also refer to the discussions on the effects of momentum contained in \cite{2987040,105555}.}
Nevertheless, at later training stages, the information filtered away by a smoothening process can be useful to further optimize the network. 
This corresponds to the fact that a coarse-grained loss would perform asymptotically in a sub-optimal fashion. 

A further, more theoretical reason in favor of switching off the entropic smoothening along the training connects to convergence. 
If the regularization is switched off completely starting from some (possibly predetermined) training step, then we can directly 
apply the convergence results of the standard stochastic gradient descent algorithm to the overall training.

Another related way to interpret the effect of an entropic smoothening is as a device enforcing an exploration/exploitation (or robustness-sensitivity) trade off. 
Progressively switching off the entropic smoothening appears to be desirable, amounting to favoring 
exploration in an early phase of the training, while enhancing the exploitation later on.
Indeed, we want to be more robust initially and then learn finer details in a subsequent training phase,
which is implemented through a \emph{``search-then-converge''} schedule \cite{2987040} for the entropic regularization.

It is interesting to draw a comparison with renormalization theory in statistical mechanics, which requires a small detour.
Renormalization in statistical physics can be described as a (systematic) procedure to filter away the information about the microscopic 
dynamics of a physical system in order to retain only the \emph{relevant} dynamics determining its low-energy or overall behavior.%
\footnote{See Appendix \ref{filter} for related comments.}
As an extreme case of renormalization, consider for instance the thermodynamic description of a gas where a small set of thermodynamic 
parameters (the temperature, the pressure and so on) accounts for the overall description of the macroscopic behavior of a microscopically very complicated system.

In a machine learning task, to some extent, one proceeds as in renormalization: one is interested in filtering away (irrelevant) information 
about the details of the dataset to keep just the relevant information needed for the task. 
Thus, seemingly, a training process should be interpretable as a dissipative flow along which irrelevant information is progressively forgot.

This intuitive picture is not sufficient to account for the training of a neural network. To understand this it is enough to note that, 
unlike the physical system (stick to the example of a gas), the neural network does not contain the microscopic information to begin with. 
In other words, the training is not simply a filtering operation, rather is it a process in which information is acquired from the dataset 
and -generically- filtered at the same time.%
\footnote{These comments are close to the information bottle-neck analysis of neural networks, see \cite{shwartzziv2017opening,michael2018on,goldfeld2019estimating}.}
In this sense, the scoping of the entropic smoothening corresponds to organizing the learning priority, coarse-grained and robust features first, finer details later on.

\section{Alternative to initialization}
\label{alternativo}

Since the entropic smoothening can provide tunable parameters which control the synaptic sensitivity,
it can be exploited to pursue an alternative solution to the so-called \emph{initialization problem} in deep networks.
Complicated neural networks, especially those that -due to their depth- entail strong hierarchies among different weights, 
can need suitable initialization procedures in order to be trained. This is the case for deep convolutional neural networks.
Although the initialization problem has been studied in detail, and suitable as well as easy initialization procedures have been devised \cite{xavier,he2015delving}, 
it is interesting to consider alternatives. In particular, one would like to seek for methods which can be applicable in 
general, independently of the specific characteristics of the architecture of the underlying neural network.
In this context, the entropic smoothening offers a viable and versatile tool. 
By considering an aggressive entropic regularization at the beginning of the training, 
one induces a strong insensitivity to the initial state.

The idea of an initially insensitive neural network, which is progressively made more sensitive during training 
matches with the arguments described in Section \ref{scoping}. Here we stress that such scoping of the entropic smoothening 
can be pushed to the extent that it makes an initialization procedure superfluous.
To substantiate this proposal we tested it in two different circumstances, both referring to image-classification tasks. 
We describe the experimental details and results in two separate subsections, Subsection \ref{MNIST} for experiments performed 
on the MNIST dataset and Subsection \ref{STL10} for experiments performed on STL10.

It is useful to stress that here we are considering an entropic regularization on the weights of
convolutional layers. This has been argued to worsen the performance of the neural network \cite{Musso:2020qmt}.%
\footnote{A similar observation applies to dropout regularization of convolutional layers 
which generically affects negatively the overall performance \cite{devries2017improved}.}
However, here we consider a regularization which is active only at an early stage of the training and which fades away completely at later stages.

Another technical observation, which applies to all the experiments described below, is that the entropic regularization 
has been enforced by taking a single extra weight configuration for each training step.
This corresponded to considering a kernel $K$ given by the characteristic function of a hypercube and approximating the convolution integral \eqref{part} by means of a (minimal) empirical sampling.
The extra configuration $\boldsymbol W'$ is sampled uniformly in a hypercubic vicinity of the original configuration $\boldsymbol W$.
More precisely, the extra sampling point $\boldsymbol W'$ is generated as a perturbation of the unperturbed configuration 
$\boldsymbol W$ according to 
\begin{equation}
 \boldsymbol W' = \boldsymbol W + \Delta \boldsymbol W\ ,
\end{equation}
where
$\Delta \boldsymbol W$ is a vector whose components $\Delta W_i$ are uniformly distributed in the interval
\begin{equation}\label{Ri}
 [-R_i,R_i]\ .
\end{equation}
The parameter $R_i$ sets the size along the $i$-th direction of the hypercube.%
\footnote{See \cite{Musso:2020qmt} for details.}
Note that $R_i$ controls the size of the smoothening vicinity, a role that in \eqref{loc_ent} 
corresponded to $\gamma^\frac{1}{2}$.

\subsection{MNIST}
\label{MNIST}

\begin{table}[ht]
 \begin{center}
  \begin{tabular}{c|c|c|c}
  layer & input channels%
  \footnote{The ``channels'' are interpreted as the height of a stack of images for the convolutional layers (\emph{e.g.} an RGB 
 color image has three channels) while it represents the total number of inputs for a fully connected layer.}
 & output channels & kernel size\\
  \hline
  conv & 1 & 1 & 3\\
  conv & 1 & 1 & 3\\
  conv & 1 & 1 & 3\\
  conv & 1 & 1 & 3\\
  \hline
  fully conn. & $20\cdot 20$ & 10 & \\
  \hline
  \end{tabular}
 \end{center}
 \caption{Deep convolutional architecture adopted for image classification on MNIST.
 }
 \label{convo1}
\end{table}
\begin{figure}
\begin{center}
 \includegraphics[width=.49 \textwidth]{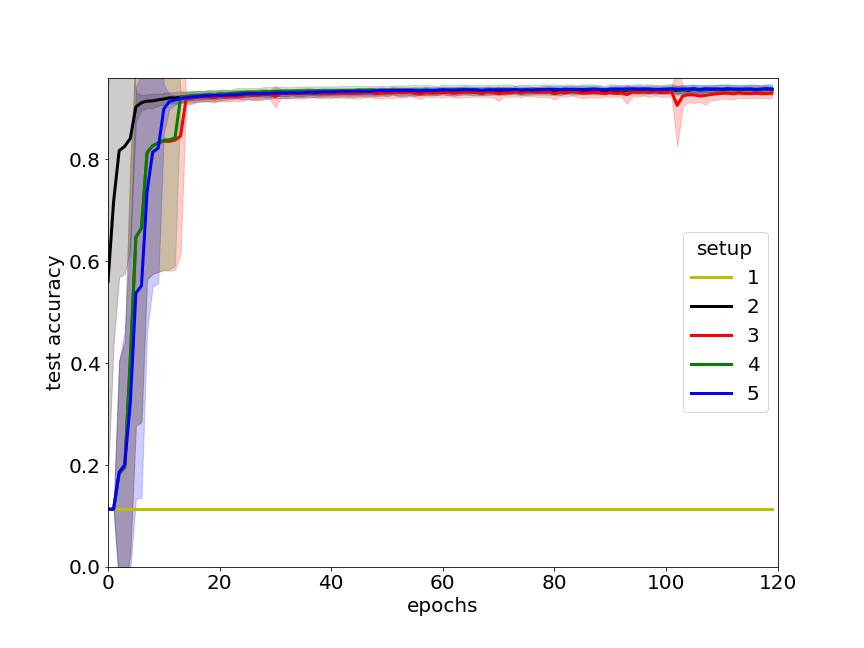}
 \includegraphics[width=.49 \textwidth]{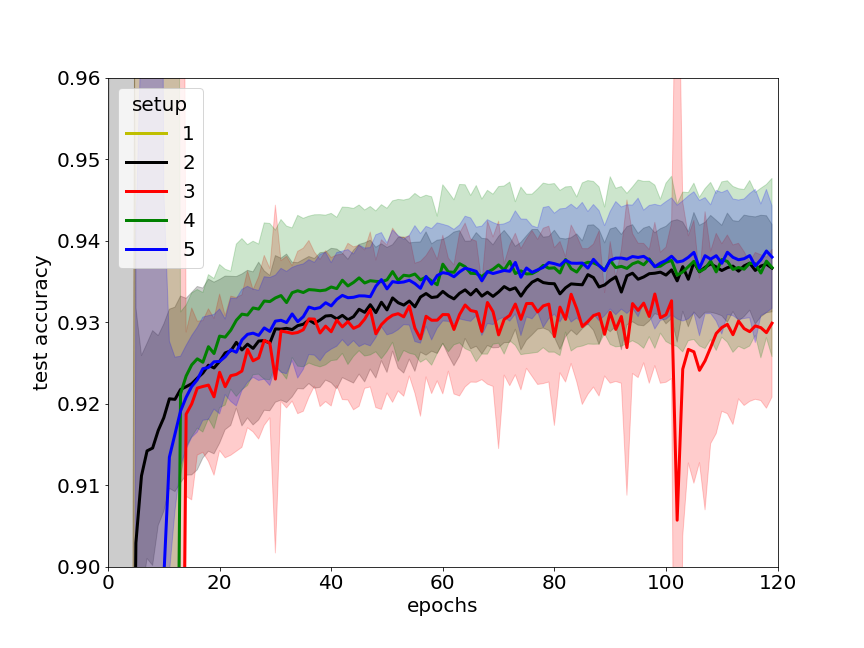}
 \caption{The two plots show the same training experiments on MNIST, but the one on the right zooms into the high-accuracy region.
 The numbering in the legend corresponds to the list of protocols described in the main text.}
 \label{MNIST_res}
 \end{center}
\end{figure}

The convolutional architecture employed in the series of experiments on MNIST is detailed in Table \ref{convo1}.
The entropic smoothening has been applied only on the weights belonging to the convolutional layers,
leaving those of the fully-connected head un-regularized. No weight decay or other sources of regularization on the weights
have been considered. The learning parameter has been kept always constant in time $\eta = 0.001$,
the same is true for the mini-batch size $C=256$ and the momentum $\mu = 0.9$. The neural network has been trained 
for 120 epochs and we have considered five distinct protocols in relation to initialization and the entropic regularization schedule,
\begin{enumerate}
 \item 
 Random initialization according to a normal distribution with zero mean and $\sigma=0.01$. No entropic regularization.
 
 \item
 Kaiming initialization \cite{he2015delving}. No entropic regularization.
 
 \item
 Random initialization according to a normal distribution with zero mean and $\sigma=0.01$. Constant, anisotropic entropic regularization
 according to $R_i = (i-1)\, \sigma$ with the index $i$ counting the layers, $i=1,...,4$.%
 \footnote{In \eqref{Ri} the index $i$ ran over the weights, here we overload the notation because we are considering that $R_i$ is equal for all the weights belonging to the same layer.}

 \item
 Random initialization according to a normal distribution with zero mean and $\sigma=0.01$. Scheduled, anisotropic entropic regularization
 starting with $R_i = (i-1)\, \sigma$, reduced by a factor $\frac{1}{3}$ for each ten training epochs (the scheduling corresponds to an exponential decay).
 
 \item
 Random initialization according to a normal distribution with zero mean and $\sigma=0.01$. Scheduled, anisotropic entropic regularization
 with $R_i(t) = \frac{i-1}{\sqrt{t}}\, \sigma$ where $t$ is a discrete time variable counting the number of training epochs elapsed.
\end{enumerate}

The results are summarized in Figure \ref{MNIST_res}. The random initialization with no entropic regularization (protocol 1) could not be 
trained. Random initialization with a constant entropic regularization (protocol 3) proved to be trainable but led to sub-optimal results. 
The remaining three protocols proved to be optimal and -essentially- equivalent. To recapitulate, a training starting from random initialization,
when performed according to a properly scheduled entropic regularization, led to equivalent results as a non-regularized case initialized 
with the Kaimimg method.

\subsection{STL10}
\label{STL10}

\begin{table}
 \begin{center}
  \begin{tabular}{c|c|c|c}
  layer & input channels & output channels & kernel size\\
  \hline
  conv & 3 & 8 & 3\\
  conv & 8 & 8 & 3\\
  \hline
  max pool & & & 2\\
  \hline
  conv & 8 & 16 & 3\\
  conv & 16 & 16 & 3\\
  conv & 16 & 16 & 3\\
  \hline
  max pool & & & 2\\
  \hline
  fully conn. & $16\cdot 20\cdot 20$ & $16\cdot 20\cdot 20$ & \\
  fully conn. & $16\cdot 20\cdot 20$ & $16\cdot 20\cdot 20$ & \\
  fully conn. & $16\cdot 20\cdot 20$ & 10 & \\
  \hline
  \end{tabular}
 \end{center}
 \caption{Deep convolutional architecture adopted for image classification on STL10.}
 \label{convo3}
\end{table}
\begin{figure}
\begin{center}
 \includegraphics[width=.49 \textwidth]{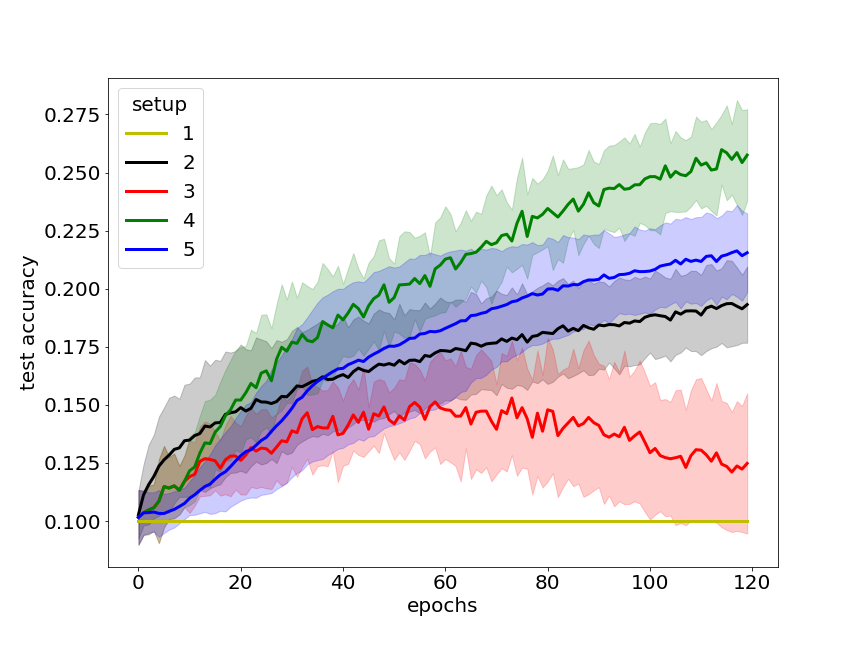}
 \includegraphics[width=.49 \textwidth]{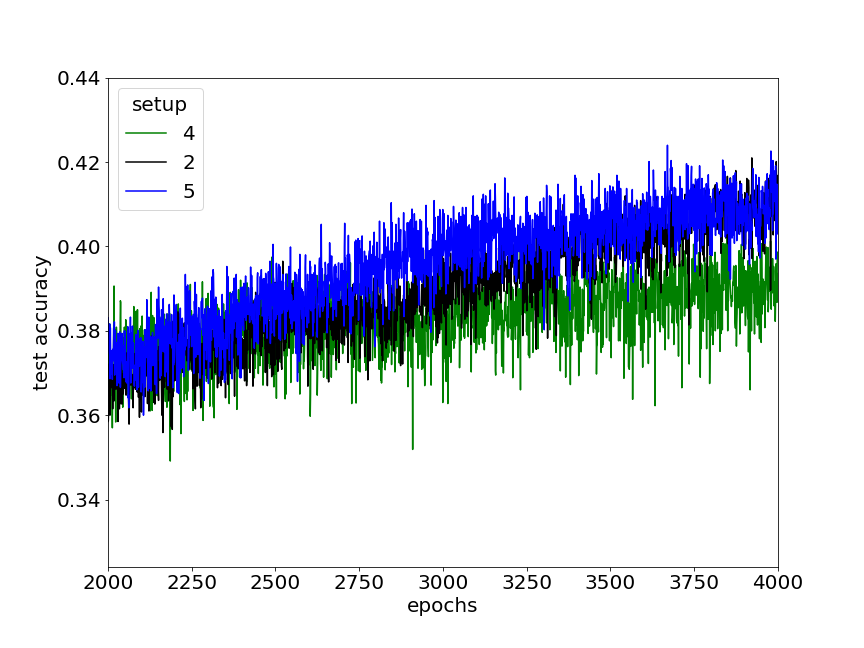}
 \caption{Training on STL10 according to the protocols detailed in Subsection \ref{MNIST} (the same as those adopted on MNIST).
 Left: early training phase. Right: large-time behavior for protocols 2, 4 and 5.}
 \label{STL10_ReLU}
 \end{center}
\end{figure}

We still consider a series of experiments similar to those described in Subsection \ref{MNIST}, 
here performed on the significantly more demanding image-classification task defined by the STL10 dataset.
To this purpose, we adopt the architecture described in Table \ref{convo3}. The network is one layer deeper than that used on MNIST,
but -apart from this difference- we consider the same experimental setups as described in the list in Subsection \ref{MNIST}.
Also regarding the training hyperparameters, we consider the same values as described there, namely, $\eta = 0.001$, $C=256$ and $\mu = 0.9$.

As it already happened for MNIST, also on STL10 the random initialization with no entropic regularization (protocol 1)
corresponds to a setup where the network does not train. The case where the entropic regularization is kept constant 
throughout the training with random initialization (protocol 3) is trainable but -again- it leads to suboptimal results. 
Actually, adopting protocol 3, a sufficiently long training can spoil completely the results obtained at an earlier stage of the training. 
Protocols 2 and 5 proved to be practically equivalent, as far as the asymptotic test accuracy is concerned, with protocol
4 reaching a slightly sub-optimal result. Nonetheless, the training dynamics is quite different among the three protocols.
A scheduled entropic regularization, especially when switched off exponentially, proved to be the best achieving protocol in 
a wide portion of the early training.

The experiments on STL10 features a richer structure with an interest on its own (especially in relation 
to early-stopping policies). They however corroborate the conclusions already reached on the experiments on MNIST: a progressively 
fading entropic regularization can provide a valid alternative to Kaiming initialization.

\section{Discussion}
\label{discu}

\subsection{Local scale invariance}
\label{ska}

We adopt the Rectified Linear Unit (ReLU) activation function for all the neurons in the network,
which is a scale covariant function, namely
\begin{equation}\label{relu_cov}
 \sigma_{\text{ReLU}}(\alpha\, x) = \alpha\, \sigma_{\text{ReLU}}(x)\ ,
\end{equation}
where $\alpha$ is a generic real constant.
The outputs $z_j$ of the network are normalized by means of a \emph{softmax} function, 
\begin{equation}\label{soft_max}
 \hat p_j = \frac{e^{-z_j}}{\sum_k e^{-z_k}}\ ,
\end{equation}
so that $\{\hat p_j\}$ can be interpreted as an empirical probability distribution over classes, in the Bayesian
sense of encoding a degree of uncertainty.

The two characteristics expressed in \eqref{relu_cov} and \eqref{soft_max} make the network prediction
independent from a scaling of all the weights of a layer by the same constant factor.
We refer to this property as \emph{local scale invariance}, where the ``local'' attribute refers 
here to depth, namely we can have a different scaling factor $\alpha_i$ for each layer $i$. Clearly, local 
scale invariance is stronger than global scale invariance, this latter corresponding to $\alpha_i=\alpha$ 
for all $i$.

If the weights of a layer are distributed with a variance $v_i$, scaling them by a factor $\alpha$
transforms the variance to $\alpha^2 v_i$. From this observation it emerges that the initialization procedures,
which tune the variance of the initial weight distribution, set a local scale (\emph{i.e.} a scale for each
layer), meaning that the initialization does not commute with a local scaling transformation. 
If we think 
to the cost function as an energy%
\footnote{To the purposes of the present discussion, one can interpret the cost function as the static potential 
of a would-be Hamiltonian system.} and to the weight configuration as the state of the system, then we have that 
local scale invariance is preserved by the energy function (\emph{i.e.} the energy is invariant under local scale transformations)
but broken by the state of the network. This corresponds to what in physics is referred to as \emph{spontaneous symmetry breaking}.

The former statement applies to the unregularized loss function ${\cal L}(\boldsymbol W)$, 
before considering the entropic regularization \eqref{loc_ent}. In fact, the regularizing term
in \eqref{loc_ent} is not invariant under local rescalings of the weights, thus neither it is so the \emph{local free loss} ${\cal F}(\boldsymbol W')$.%
\footnote{To avoid confusion, we remind ourselves that the adjective \emph{local} in \emph{local free loss} refers to locality in the synaptic space.}
This situation is referred to as \emph{explicit breaking} of the local scaling symmetry. Indeed, one can think to 
\begin{equation}
 {\cal L}_{\gamma} (\boldsymbol W, \boldsymbol W') = {\cal L} (\boldsymbol W) + \frac{\gamma}{2} \|\boldsymbol W - \boldsymbol W'\|_2^2\ ,
\end{equation}
appearing at the exponent in the integrand of \eqref{loc_ent}, as a modified energy function, where $\gamma$ is a source of explicit symmetry breaking.
In other words, $\gamma$ introduces explicitly a scale into the formerly scale-invariant problem. In the case in which we consider different 
$\gamma$'s for different directions in the synaptic space, we introduce more than one explicit scale into the problem.

It is relevant to note that these statements about local scaling symmetry apply also when considering the training dynamics.
Specifically, if the loss is invariant under local scalings, then the training step (\emph{i.e.} the gradient descent step) 
commutes with a local rescaling. This means that one can perform the rescaling and the training step in either order and reach 
the same final state. This is due to the linearity of the gradient descent algorithm and would cease to apply when considering higher-order 
descent algorithms.

Explicit breaking of the local scale invariance, either implemented by means of a local-entropic loss function or by considering 
activation functions which are not scale covariant, can be contrasted with the so-called \emph{natural gradients} approach \cite{Amari1996NeuralLI,pascanu2014revisiting}. 
This latter employs a normalization technique, based on approximated information-theoretic arguments, to avoid the scaling ambiguity.
More recently, normalization techniques aimed at removing the local scaling ambiguity have been studied in \cite{poggio2019theoretical,liao2020generalization}.

Although the local scale invariance is in many respects similar to a redundancy which we have to project away%
\footnote{In theoretical physics, one would speak of a gauge invariance which needs to be fixed.}, the considerations about its breaking, either spontaneous or explicit, might have a relevant role. 
Specifically, one could consider how the interplay between the spontaneous scale introduced by initialization 
and the explicit scale introduced by local entropic regularization affect the network training and eventual performance.

There are at least four further observations which connect to local scale invariance:
\begin{itemize}
 \item Scale invariance emerges in common but very specific setups, for instance, 
 in networks characterized by the exclusive adoption of ReLU activations. 
 These systems can thus be regarded as a fine-tuned family within the wider set of possible networks that are not scale invariant. 
 Apart from their practical interest \cite{prop}, scale-covariant activations are simpler to study. As such, they can constitute a good starting 
 point in view of studying how the scales introduced by non-covariant activations may affect the behavior of the networks.%
\footnote{We leave a systematic experimental study of these aspects to the future, especially because the theoretical picture 
is likely more involved (and probably more interesting, too). Nonetheless, we give some further comments on this in Subsection \ref{acti}.}
 
 \item The regularized loss function ${\cal F}(\boldsymbol W)$ introduced in \eqref{loc_ent} is defined in terms of a convolution integral 
 over the synaptic space. Although it introduces into the problem an explicit scale through $\gamma$, 
 this does not \emph{lift} the flat directions of the original loss associated to local scale invariance. 
 Equivalently, both the original loss ${\cal L}(\boldsymbol W)$ and the regularized one, ${\cal F}(\boldsymbol W)$, 
 have a Fisher matrix with some zero eigenvalues associated to the local scaling freedom.
 
 \item As already commented in Section \ref{scoping}, complexity theory and, specifically, the concept 
 of \emph{effective capacity} \cite{berezniuk2020scaledependent} suggest that also the dataset can induce 
 the notion of a scale into the synaptic space. 
 This corresponds intuitively to the idea that a richer dataset defines a more detailed loss.%
 \footnote{It is relevant to observe that fixing the local scale invariance by means of normalizing
 the weights of each layer defines a sphere for each subspace (of the synaptic space) corresponding to a layer.
 In other words, if $\boldsymbol W_i$ is a vector whose components represent the weights of the $i$-th layer,
 its normalization means that $\boldsymbol W_i$ spans a spherical surface. Such surface is compact,
 and compactness is crucial to argue that a finite dataset can induce a physical scale in the synaptic space. 
 }
 \item Individuating the physical scales and the hierarchies among them is at the basis of 
 possible effective descriptions of neural networks \cite{Cooper1973,Gabri2020,phdthesis}. Scale transformations generate the renormalization group whose fixed points 
 are associated to criticality, a condition, or regime, which could radically simplify the analysis and the training 
 of the networks \cite{schoenholz2017deep}. Entropic probes have been considered recently in this context, see \cite{erdmenger2021quantifying}.
\end{itemize}

\subsection{Related frameworks}
\label{rel_fra}
 
In practice, the computation of the convolution integral \eqref{loc_ent} is not convenient as it would entail a significant extra cost.
Rather, one considers an empirical proxy for such an integral obtained by sampling a discrete set of points according to a suitable 
sampling criterion.
A similar approach is considered in \emph{entropic least action learning}, where a number of real copies of the original system are trained 
concomitantly and coupled to one another by means of an attractive interaction, whose intensity is typically increased along training \cite{Baldassi_2016,Baldassi_2019}. 
Such \emph{elastic regularization} implements a soft version of a distance constraint between the machines working in parallel
\cite{zhang2015deep,zhang2016distributed}.

Another framework exploiting parallelism to enforce an entropic bias and, more generically, to enhance the exploratory character of an algorithm
is quantum annealing (see for instance \cite{Venegas_Andraca_2018}). In quantum annealing the support of the wave-function corresponds to the explored region. In such a framework, an external control enhancing the focusing of the wave-function 
along the training would -at least intuitively- parallel the increasing binding interaction among parallel machines in least-action learning.

\subsection{Minimal extra cost}
\label{opt}

As already mentioned, local entropic regularizations can be approximated by suitable sampling techniques.
The sampling, although being in general cheaper than the actual computation of the integral,
amounts to extra evaluations of the loss function and computation of its gradient. Clearly, this correspond to 
an increased computational cost at each training step. Two relevant comments in this respect are the following.
\begin{itemize}
\item 
Numerical experiments show that the cost can be minimal, yet still leading to useful effects. In particular, already adding 
just one extra sampling point in the vicinity of the original point yields the positive effects of local entropic regularization.%
\footnote{A related observation was already made in \cite{Musso:2020qmt}, where experiments performed with a 
bi-layer fully-connected network on the Fashion-MNIST image-classification task yield very similar results
irrespective of the fact that the partial entropic regularization relied on either 5 or 9 sampling points.
Note that these numbers are very small with respect to the dimensionality of the problem (the number of weights).
For a connection between the number of sampling points and the Parisi parameter $m$ see \cite{prop}.}  
\item
The extra samplings do not require re-loading the image mini-batch, which is in general a costly operation.
This represents the main difference between the approach described here from that usually adopted in 
entropic least action learning (see Subsection \ref{rel_fra}) where each machine working in parallel
is provided with a different mini-batch of training samples.
\end{itemize}

\subsection{Gradient de-noising}
\label{deno}

 A local entropy regularization extracts more information at each training step
 by sampling the point and its vicinity. As such, it has a \emph{de-noising} effect on the 
 computed gradient. Momentum, too, has a denoising effect on the training gradient.

 Momentum integrates information coming from previous training steps. It does not entail 
 an increase of training cost, but only in memory resources (which should be generically cheap).
 Conversely, partial local entropy exploits spatially and temporally local information.
 It requires an additional sampling cost (see Subsection \ref{opt}). Momentum is ``conservative'', exploiting information coming from the past and resisting 
 to update it by means of an inertial update rule; partial local entropy, instead,
 exploits as much as possible the current circumstance independently of how one has 
 reached it. The two things might not be as independent as they  
 seem at first: the current weight configuration is conditioned by previous training so, even if 
 we decide to forget the gradient information collected in the previous steps,
 there is some information correlated to it encoded in the current state. Yet, this would be very difficult to disentangle explicitly.

\subsection{Size and shape of the solution clusters}
\label{shape}

Stochastic gradient descent and its regularized versions typically encounter solutions belonging to clusters of configurations leading to similar test accuracy. 
The size of the solution cluster, that is, the Gardner volume, can be assessed with entropic probes,
either analytically (when treatable) or numerically \cite{refId0,prop}. In a setup where the entropic regularization
is non-trivial at the end of the training, the final $\gamma$ encodes a bias on the size of the encountered cluster of
solutions. Therefore, a suitably anisotropic $\gamma$ can be used as a probe of the shape
of the solution cluster \cite{Musso:2020qmt}. In this context, we should recall that -if working with 
a scale invariant network- the concept of shape is meaningful only after a suitable normalization scheme 
has been adopted. For instance, normalizing all the layers to a fixed and pre-determined value.
This latter observation represents a refinement on the critical observations about the concept of ``wide valleys'' discussed in \cite{poggio2019theoretical}.

\subsection{Activation functions which are not scale covariant}
\label{acti}
 
\begin{figure}
 \begin{center}
  \includegraphics[width=.49\textwidth]{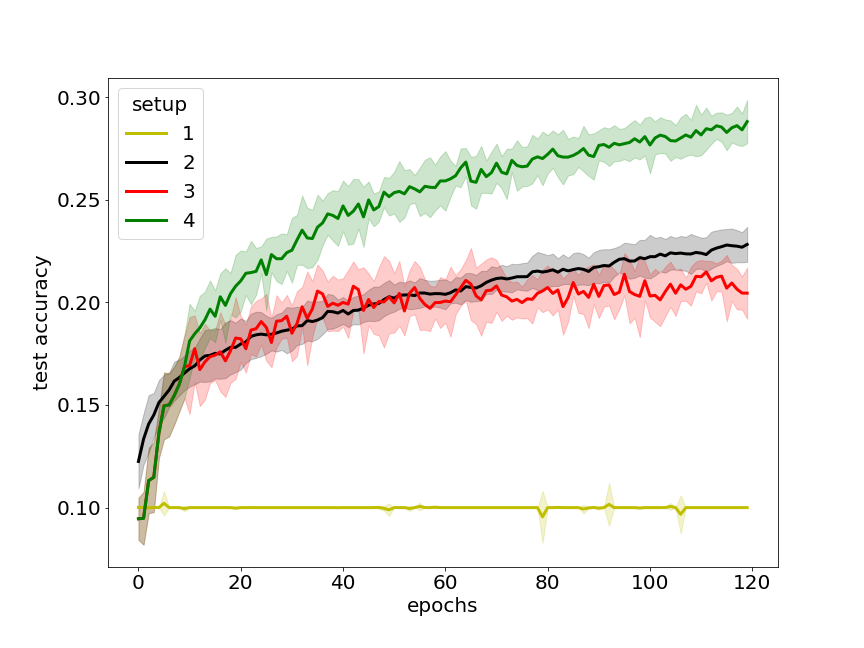}
 \end{center}
 \caption{Test accuracy during training for a convolutional neural network specified in Table \ref{convo3}, 
 but with $\tanh$ activations instead of ReLU. The task, image classification on STL10, and the hyperparameters
 are the same as those described in Subsection \ref{STL10}.}
 \label{STL10_tanh}
\end{figure}
The activation functions can have a relevant impact on the properties of the loss landscape.
We have already seen an explicit example of this when discussing scale invariance in Subsection \ref{ska} and, specifically,
when observing that the scale covariance of ReLU activations produces some exactly flat directions in the loss landscape.
As shown both analytically and numerically in \cite{prop}, the robustness properties of the solution 
clusters may depend on the choice of activation functions. 

Considering a non-scale covariant activation function, namely $\tanh$, lifts the flat directions
associated to the local scale transformations discussed in Subsection \ref{ska}. In particular,
the ratio between the intrinsic scale dictated by the $\tanh$ activation (related to the steepness of 
the transition between the two asymptotic saturating behaviors) and the scale (or scales, in the anisotropic case)
enforced by $\gamma$ could produce some observable effects in the training dynamics. 
Yet, some preliminary experiments -performed on STL10 with an identical setup as the one described in Subsection \ref{STL10}
where ReLUs have been substituted with $\tanh$ activations- show a training behavior which is qualitatively analogous.
Compare Figures \ref{STL10_tanh} and \ref{STL10_ReLU}.

\subsection{Dynamical landscape}

Suppose that the training dynamics has some time dependence on top of the scoping of the $\gamma$ parameters or a scheduling of the learning rate.
Namely, some time dependence which is related to the task evolution in time. An explicit example of this can be provided
by a dataset which changes over time. If the task evolution is rapid enough, we can think that the training dynamics 
finds itself constantly in a situation similar to being in an early training stage.
As argued in Section \ref{alternativo}, this is the circumstance in which an entropic regularization can provide an advantage, 
for convolutional weights too. Thus, we can think that $\gamma(t)$ should depend on time in response to the variations of the dataset.
Roughly, we need a higher $\gamma$ when the dataset variations are stronger, while we need a decaying $\gamma$ when the dataset 
is static or quasi-static.
To rephrase, $\gamma(t)$ seems to need to be adaptive and related to the time-dependent properties of the task.%
\footnote{In this sense, the initial condition of a training with a constant dataset can be interpreted as an abrupt change, 
or a quench, separating the time before training from the training time.
As such, the early training profits from a strong entropic regularization as a response to the initial conditions.}
In physical terms, $\gamma(t)$ should adequately respond to the external driving.
Entropic regularization could therefore be interesting in the context of the problems related to catastrophic forgetting \cite{cata}.

\subsection{Final remarks}

We showed that an anisotropic and suitably scheduled entropic regularization can provide a tunable alternative to initialization procedures
for deep neural networks. Moreover, the entropic protocols can enhance the test-accuracy, especially in early training phases. The training 
dynamics featured by the entropic protocols appears to be very rich and sensitive to both the architecture and hyper-parameters, on the one side,
and to the characteristics of the task, on the other. As such, a systematic characterization of the effects of anisotropic and scheduled 
entropic regularization is a very complex and wide task. We provided here a first exploratory round of experiments to stress the potential 
and the theoretical interest. An in-depth experimental study constitute however a promising route for future investigations, 
which could be performed with either an exploratory or an exploitative attitude, that is to say, it can be pursued with a tunable 
inclination towards practical applications.

\section{Acknowledgements}

A special acknowledgment goes to Manuel Fernández Delgado and Giorgio Musso for interchanges and feedback on the draft.

I would like also to thank Amparo Alonso Betanzos, Antonio Amariti, Carlo Baldassi, Brais Cancela Barizo, Xabier Cid Vidal, Aldo Cotrone, Thomas Dent, Carlos Eiras Franco, 
Andrés Gómez Tato, Carlos Hoyos, Alessandro Ingrosso, Estelle Maeva Inack, Esteban Requeijo Gonzalez, Lorenzo Rosasco, Silvia Villa and Riccardo Zecchina
for stimulating and interesting conversations.

\appendix 

\section{Entropic smoothening as a filtering process}
\label{filter}

Consider the definition of the local partition function \eqref{part} as a \emph{filtering} procedure applied to the Boltzmann weight $e^{-\beta{\cal L}(\boldsymbol W)}$ and expressed mathematically through the convolution with the integration kernel $K$.

To gain intuition about this, it is useful to think of two extreme cases. First, take the trivial kernel
\begin{equation}
 K(\boldsymbol W, \boldsymbol W') = 1\ .
\end{equation}
In this case the filtering due to \eqref{part} reduces to the simple integral of the Boltzmann weight over the configuration space. This is the standard definition of the partition function in statistical mechanics. In a learning perspective, it corresponds to having filtered away as much information as possible, just retaining the thermodynamic information actually encoded in $Z$.

An opposite circumstance occurs when the integration kernel takes the form of a Dirac delta,
\begin{equation}
 K(\boldsymbol W, \boldsymbol W') = \delta(\boldsymbol W, \boldsymbol W')\ .
\end{equation}
In such a case, the convolution \eqref{part} simply returns the original Boltzmann weight, the delta corresponding to an identity operator. Thus, no information has been integrated away by the convolution.

The cases with generic kernels $K$, as well as the specific examples considered in the main text (like that of a Gaussian $K$), fall intuitively between the two extreme cases just described. 
In other terms, partial local entropy can be interpreted as a procedure which refines the thermodynamic description, 
but still filters some microscopic information away. Specifically, local thermodynamic functionals defined by means of a convolution with a 
kernel with compact support (or, at least, a kernel which is significantly different from zero on a compact region of the synaptic space)
represent essentially the local contribution to the associated thermodynamic potential. For instance, \eqref{loc_ent} represents the local free energy contribution to the standard Helmholtz free energy

\section{Physically motivated pruning of a perceptron leads to a convolutional network}

\begin{figure}[h!]
 \begin{center}
  \includegraphics[width=0.28 \textwidth]{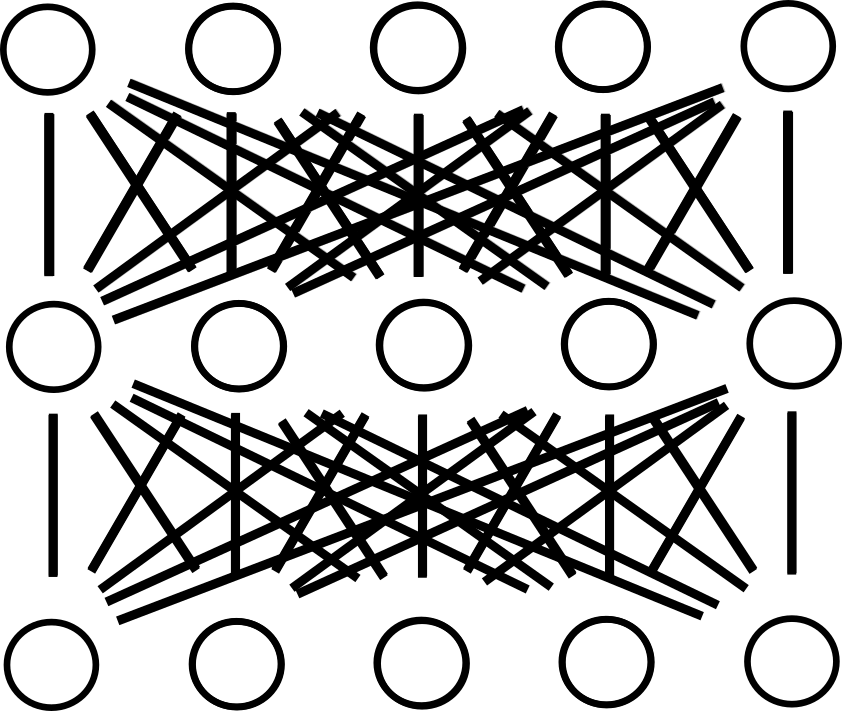}\hspace{20pt}
  \includegraphics[width=0.28 \textwidth]{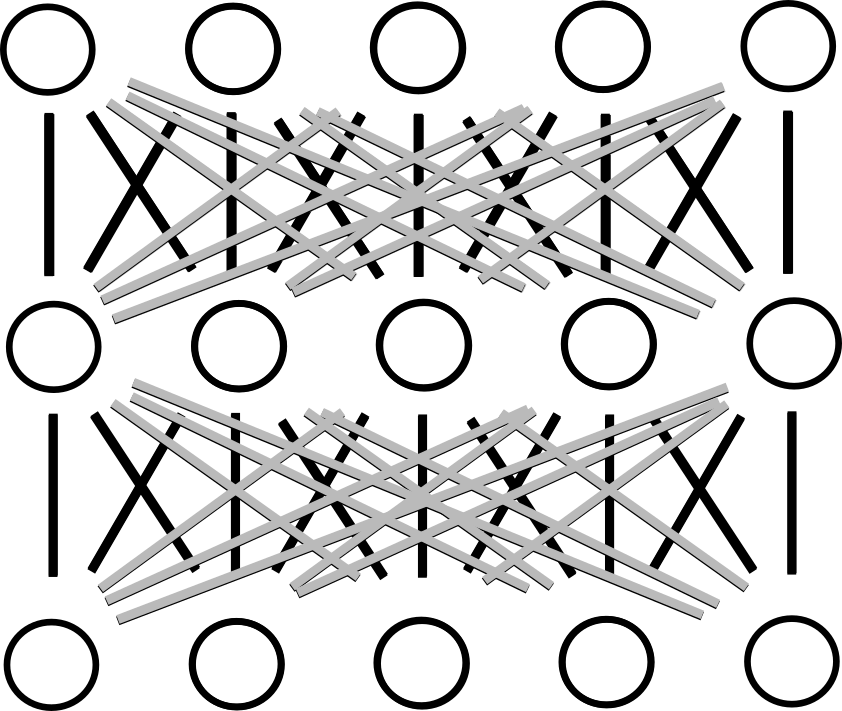}\hspace{20pt} 
  \includegraphics[width=0.28 \textwidth]{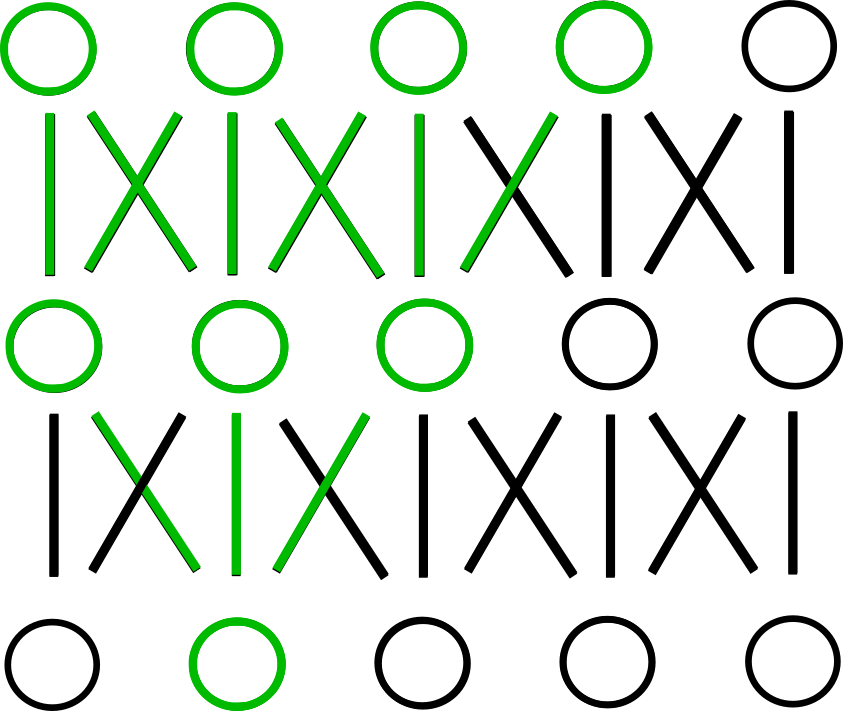}\\ \vspace{20pt}
  \includegraphics[width=0.28 \textwidth]{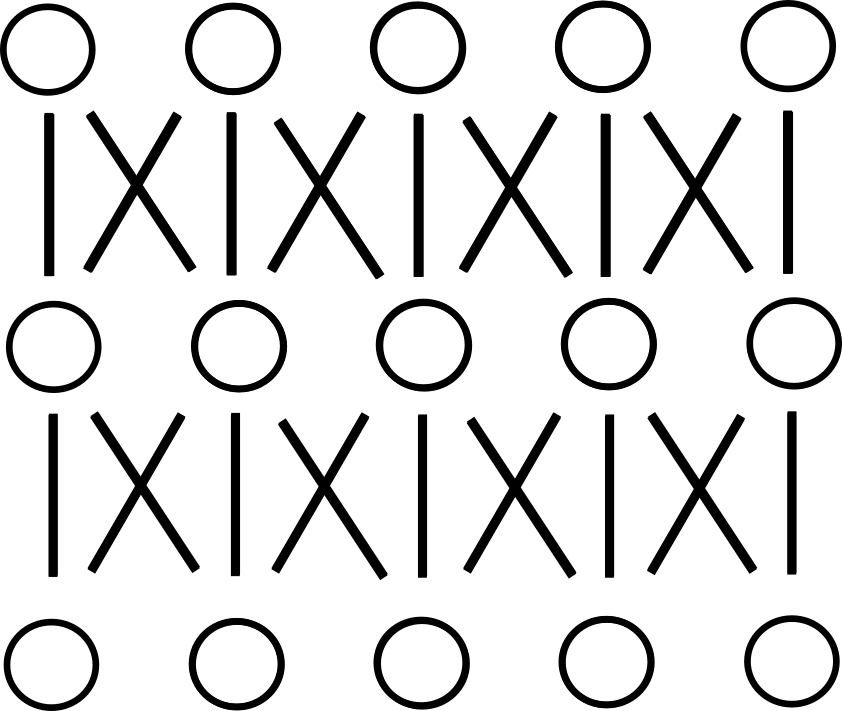}\hspace{20pt}
  \includegraphics[width=0.28 \textwidth]{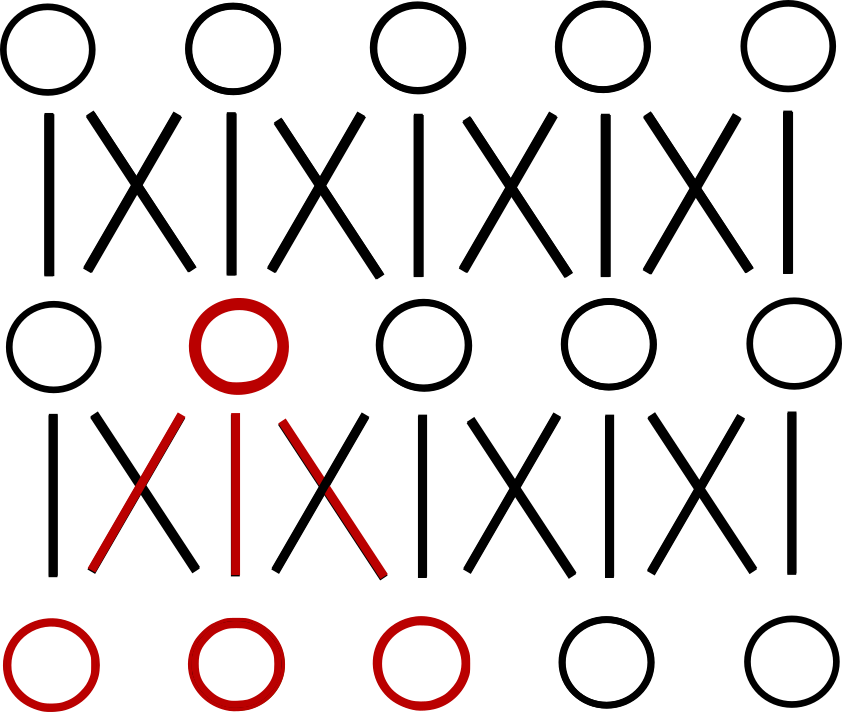}\hspace{20pt}
  \includegraphics[width=0.28 \textwidth]{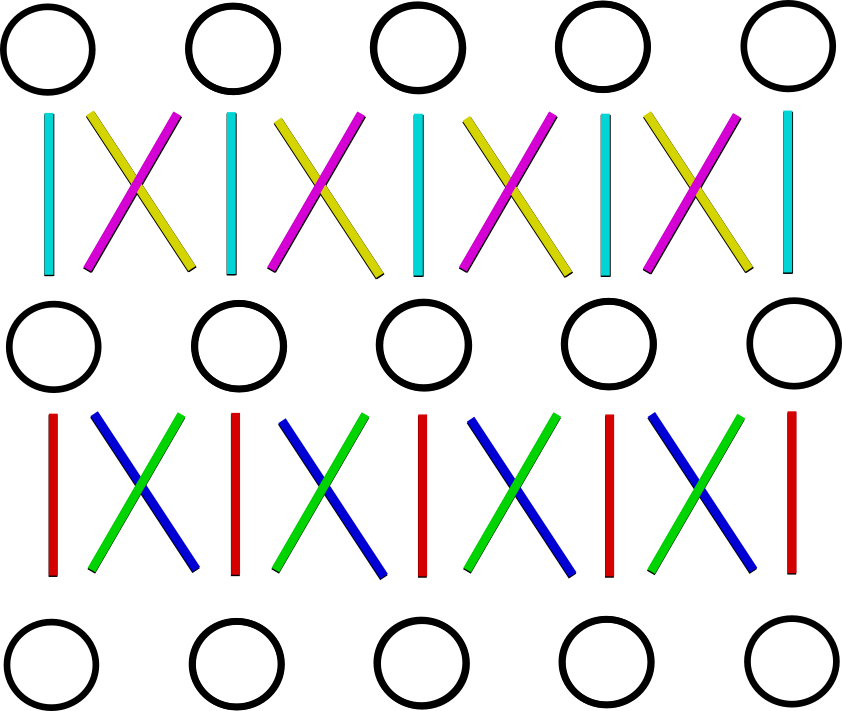}
 \end{center}
 \caption{Convolutional architecture as a physically-inspired pruning of a fully-convolutional network.}
 \label{conv}
\end{figure}

In the main text, the inspiration coming from physics has been frequently appealed. Still resorting to physics, one can motivate the derivation of a convolutional architecture from a fully-connected one on the basis of generic principles. Specifically, locality and translational covariance.

This not only provides an organizing principle to ``expect'' that convolutional architectures may be convenient, it also provides a suggestive connection between the network and the geometry of spacetimes endowed with a light-cone structure.

Let us clarify by means of an explicit example. Consider a small fully-connected network formed by 15 neurons. For simplicity, take a cylindrical network, and consider three layers counting five neurons each, see Figure \ref{conv}. Although having just 15 neurons, this multi-layer perceptron in its fully-connected configuration presents an already quite complicated link structure.
It seems thereby natural to devise some simplified version, namely to find some criterion to reduce the number of links. This is sometimes referred to as a \emph{pruning} procedure.

Suppose that the input layer is on the bottom of the pictures, so that the depth of the network coincides with height in the pictures of Figure \ref{conv}. 
Also, associate the vertical direction (\emph{i.e.} still the depth) to time, according to the logic that information flows from the input toward the output of the network. 
Now, assume that the ``signals'' have a finite horizontal propagation speed. This means, for instance, that the output of a neuron $A$ can reach only the neurons that 
are sufficiently close to the neuron lying on top of $A$, one layer deeper. Notice that this statement is introducing a specific notion of \emph{locality} into the network. 
Actually, we are inducing a structure of light-cones, see Figure \ref{conv}. Specifically, a neuron in the input layer can affect only those neurons which belong to a conical 
region above it, progressively widening when going deeper (\emph{i.e.} as time progresses). Thus, the neurons belonging to this \emph{future light cone} are the only ones 
which can be causally connected to the neuron sitting at the vertex of the cone.

Apart from the light-cone interpretation, note that locality has motivated a pruning technique of the fully-connected network which directly returned a ``convolutional'' structure. Actually, we are still half-way on our path to a convolutional architecture. To reach there we still need to comment translational symmetry.

Let us first observe that we can define also a \emph{past light cone}. Namely, a neuron sitting at a point within the network can be influenced by all 
neurons which belong to a cone, progressively widening towards the input, whose vertex is the original neuron position. With different words, we have just re-expressed what is usually referred to as the neuron's \emph{receptive field}.

The last, essential, ingredient to reach the convolutional network is related to symmetry, specifically, covariance of the network with respect to translational symmetry. 
Still referring to Figure \ref{conv}, consider a pruned network where the weights connected by a horizontal (discrete) translation are constrained to be the same. The color coding in Figure \ref{conv} is meant to illustrate this idea.
Now, combining the observation about the structure of the receptive fields and translational covariance, once can directly prove that the operation encoded in the neural network is actually a convolution, in the standard sense.

\bibliography{DB} 
\bibliographystyle{ieeetr}

\end{document}